\documentclass[acmsmall]{acmart}

\usepackage{url}
\usepackage{graphicx}
\usepackage{acronym}
\usepackage{mdframed}
\usepackage{framed}
\usepackage{balance}
\usepackage{hyperref}
\usepackage{mathptmx}
\usepackage{amsmath}
\usepackage{color,soul}
\usepackage{makecell}
\usepackage{caption}
\usepackage{subcaption}
\usepackage{fixltx2e}
\usepackage{stfloats}

\AtBeginDocument{%
  \providecommand\BibTeX{{%
    \normalfont B\kern-0.5em{\scshape i\kern-0.25em b}\kern-0.8em\TeX}}}

\setcopyright{rightsretained}
\acmJournal{PACMHCI}
\acmYear{2023} \acmVolume{7} \acmNumber{CSCW2} \acmArticle{330} \acmMonth{10} \acmPrice{}\acmDOI{10.1145/3610179}
\received{January 2023}
\received[revised]{April 2023}
\received[accepted]{May 2023}
\begin{document}

\title[Mapping Technology-mediated Harms, Risks and Vulnerabilities]{Co-creating a Transdisciplinary Map of Technology-mediated Harms, Risks and Vulnerabilities: Challenges, Ambivalences and Opportunities}

\author{Andrés Domínguez Hernández}
\orcid{0000-0001-7492-7923}
\affiliation{%
  \institution{University of Bristol}
  %\streetaddress{}
  \city{Bristol}
  %\state{}
  \country{UK}
  %\postcode{}
}
\email{andres.dominguez@bristol.ac.uk}
\author{Kopo M. Ramokapane}
\orcid{0000-0001-8420-3929}
\affiliation{%
  \institution{University of Bristol}
  %\streetaddress{}
  \city{Bristol}
  %\state{}
  \country{UK}
  %\postcode{}
}
\email{marvin.ramokapane@bristol.ac.uk}

\author{Partha Das Chowdhury}
\orcid{0000-0002-5367-6659}
\affiliation{%
  \institution{University of Bristol}
  %\streetaddress{}
  \city{Bristol}
  %\state{}
  \country{UK}
  %\postcode{}
}
\email{partha.daschowdhury@bristol.ac.uk}

\author{Ola Michalec}
\orcid{0000-0003-3807-0197}
\affiliation{%
  \institution{University of Bristol}
  %\streetaddress{}
  \city{Bristol}
  %\state{}
  \country{UK}
  %\postcode{}
}
\email{ola.michalec@bristol.ac.uk}

\author{Emily Johnstone}
\orcid{0009-0002-7509-5174}
\affiliation{%
  \institution{University of Bath}
  %\streetaddress{}
  \city{Bath}
  %\state{}
  \country{UK}
  %\postcode{}
}
\email{ekj27@bath.ac.uk}

\author{Emily Godwin}
\orcid{0009-0000-2189-7847}
\affiliation{%
  \institution{University of Bath}
  %\streetaddress{}
  \city{Bath}
  %\state{}
  \country{UK}
  %\postcode{}
}
\email{eg780@bath.ac.uk}

\author{Alicia G Cork}
\orcid{0000-0003-2892-9615}
\affiliation{%
  \institution{University of Bath}
  %\streetaddress{}
  \city{Bath}
  %\state{}
  \country{UK}
  %\postcode{}
}
\email{ac974@bath.ac.uk}

\author{Awais Rashid}
\orcid{0000-0002-0109-1341}
\affiliation{%
  \institution{University of Bristol}
  %\streetaddress{}
  \city{Bristol}
  %\state{}
  \country{UK}
  %\postcode{}
}
\email{awais.rashid@bristol.ac.uk}

\renewcommand{\shortauthors}{Andrés Domínguez Hernández et al.}

\begin{abstract}
The phrase ``online harms’’ has emerged in recent years out of a growing political willingness to address the ethical and social issues associated with the use of the Internet and digital technology at large. The broad landscape that surrounds online harms gathers a multitude of disciplinary, sectoral and organizational efforts while raising myriad challenges and opportunities for the crossing entrenched boundaries. In this paper we draw lessons from a journey of co-creating a transdisciplinary knowledge infrastructure within a large research initiative animated by the online harms agenda. We begin with a reflection of the implications of mapping, taxonomizing and constructing knowledge infrastructures and a brief review of how online harm and adjacent themes have been theorized and classified in the literature to date. Grounded on our own experience of co-creating a map of online harms, we then argue that the map---and the process of mapping---perform three mutually constitutive functions, acting simultaneously as method, medium and provocation. We draw lessons from how an open-ended approach to mapping, despite not guaranteeing consensus, can foster productive debate and collaboration in ethically and politically fraught areas of research. We end with a call for CSCW research to surface and engage with the multiple temporalities, social lives and political sensibilities of knowledge infrastructures.

\end{abstract}





\maketitle

\section{Introduction}
Recent years have seen growing debate among governments, academia, and civil society around a host of safety and ethical issues associated with the ubiquity, scale, and speed afforded by digital technologies. Some of these pertain to the widespread use of online forums and social media platforms, including the rise of mis/disinformation of various kinds, the spread of hate speech and toxic content, cyberbullying, online harassment, and other types of abuse of vulnerable groups including children~\cite{dominguez_hernandez_ethical_2023}. Others stem from people’s everyday interactions with a variety of digital infrastructures and data-driven services where ethical issues manifest in injustices caused by automated decision-making; misuse, extraction, and  exploitation of people’s personal data; or the ever more pervasive forms of surveillance impinging on people’s freedoms~\cite{birhane_algorithmic_2021}. Investigations in these areas are led by a wide diversity of researchers across cybersecurity, data science, computer science, criminology, psychology, media and communication studies, philosophy, human-computer interaction, science and technology studies, law, among others. Much of this work has aimed at understanding the negative impacts of digital technology in society as well as developing tools to detect, predict and mitigate harmful outcomes. 

In the last 5 years, there have been more concerted governmental efforts in Europe, such as the EU's proposed regulations on platforms\cite{EUcom2022} and artificial intelligence \cite{european_parliament_eu_2023}, and the UK's ``Online Safety'' bill \cite{DCMS2022safety}, which signal a willingness to deal with the global scale challenges posed by big data and social media, reign in the power of large technology companies, and regulate the digital economy~\cite{gorwa_state_2020, royal_society_online_2019, dcms_research_2019, tufekci_algorithmic_2015}. These efforts have influenced the funding of academic research directed at tackling the most pressing individual and social harms through more evidence and tools to inform legislation, law enforcement, oversight and regulation. At the same time, research funding institutions increasingly view crossing disciplinary boundaries as an imperative for dealing with the biggest challenges of contemporary digital societies. While this reflects consistently in research funding calls and agendas, inter/trans-disciplinary collaboration is known to be challenged by entrenched academic cultures, hierarchies of knowledge, and prevailing institutional and power structures~\cite{felt_response-able_2017}. These issues are particularly salient in the highly complex and emerging landscape of ``online harms'' which is open to a diversity of conceptual definitions, terminologies, disciplinary orientations and political agendas.  

Research in CSCW has held a longstanding interest in studying how different designs, visualizations, and modalities of knowledge infrastructures support knowledge exchange and scientific collaboration~\cite{karasti_enriching_2006, young_redefining_2015}. In this paper, we wish to build upon and contribute to this body of work by drawing on our experience of co-creating a collaborative tool aimed at mapping and visualizing the vast area of research around online harm. 
We report on our work as academics within the UK National Research Centre on Privacy, Harm Reduction and Adversarial Influence Online -- REPHRAIN (hereafter the Center),\footnote{https://www.rephrain.ac.uk/about/} which was founded with a remit around ``protecting citizens online''. The Center funds a wide range of theoretical, empirical, and experimental projects from various areas of research and disciplines. While each of these projects has its own timelines and deliverables, the Center encourages collaboration through the funding of cross-cutting work and spaces for co-creation among a cohort of over 100 affiliated investigators and external partners. To that end, one of the core aims of the Center was to co-create a digital knowledge resource---known as the ``map of online harms''---that would provide relevant, up-to-date material to the Center’s stakeholders in academia, policy, industry, and third-sector organizations. 

The goal of this paper is to analyze and draw lessons from the co-creation of a collaborative digital artifact within a highly complex and contentious and evolving area of research where diverse disciplines meet. In particular, we bring attention to how an open-ended, always in the making, approach to co-creation can be generative in different ways to collaborative endeavors where consensus might be difficult to achieve. We argue that the map as an artifact---and the process of mapping itself---perform three interlocked functions for scientific collaboration and knowledge exchange, simultaneously acting as \emph{method}, as \emph{medium}, and as \emph{provocation}. 

In the first part of the paper, we look at the implications of the practices of classification, mapping, and taxonomizing in settings where different epistemic communities (seek to) coalesce. We then review how different disciplines have defined, theorized, categorized, and synthesized evidence around the broad arena of online harms. In the second part, we elaborate each of the three map functions grounding the analysis on our 18-month journey of co-creating an interactive map of online harms and developing a framework of classification and visualization in collaboration with the Center stakeholders. We discuss the challenges encountered along the way throughout a series of co-creative moments including scoping sessions, data curation, language negotiation, visualization, and maintenance. We conclude by discussing how an open-ended approach to knowledge mapping, despite not guaranteeing consensus, can foster debate and collaboration in ethically and politically fraught areas of research. 

% 4. Structurally, there is a mismatch between the motivation of the study and the stated contribution. While the motivation of the study centers around the goal of taxonomizing online harm, the written contribution focuses on the potential of mapping as a method. The author might want to consider which of the two they want to focus on. These could also very well be separated into two full papers, as they represent very different lines of contributions (empirical contribution through co-designing versus methodological contribution through novel approaches).

\section{Mapping practices and their implications}
\subsection{When the map is not the territory}
%\subsection{Classification and its consequences}

Studying the implications of classifying and visualizing knowledge has been a long-standing area of interest across Computer-Supported Cooperative Work (CSCW), Science and Technology Studies (STS), and Human Geography. Attuned to the power dynamics present in collaboration during knowledge production, these disciplines reveal the assumptions and agenda inscribed in maps or taxonomies \cite{bowker2000sorting, crampton2018introduction}. Sociotechnical studies of classification and visualization also demonstrate how the mundane work of standardization lays the ground for the creation of knowledge infrastructures, which then influences how people working together operate in society---for example, a diagnostic manual helping medical professionals decide between two similar health conditions, or a national research impact framework encouraging researchers to publish within a particular discipline \cite{mollers2017mundane, bowker2000sorting, monteiro2014innovation}. Historical and ethnographic studies of scientific production have demonstrated that the act of classifying is not a neutral process of reflecting a ``natural order’’ of the world. Instead, classifications are necessarily shaped by the goals of those who create them\cite{mattern2015mission}. Classifying is in essence an act of sorting out, highlighting the existence of certain things at the expense of others~\cite{bowker2000sorting}.

Efforts to classify, systematize, and accredit knowledge are characterized by their long history spanning multiple disciplines and professions. \footnote{The outputs range from the early examples of encyclopedias in Ancient Rome \cite{Britannica2023}, 18th Century Linnaeus’ taxonomy of species\cite{frangsmyr1983linnaeus}, to Bodies of Knowledge in contemporary professions like IT or Civil Engineering \cite{stinson1995body}} Many present-day classifications have become so widely accepted that they rarely get questioned in public debates, be it the metric system, diagnostic criteria for health conditions or spelling conventions \cite{stoytcheva2013boundary}. 

The more standards are associated with authorities and expert gate-keeping, the more they are prone to resistance that views them as ``imperialist imposition of representation, coercion, silencing, and fragmentation’’ [p.413]\cite{star1989institutional}. 
A famous remark ``the map is not the territory’’ by Alfred Korzybski \cite{korzybski1998map} points to this complex relationship between reality and its representations. The social acceptance of maps, graphs, bibliometrics and other scientific visualizations typically rely on the authority (i.e. power) of credentialed scientists and universities to tell stories with data. However, these stories are always selective, partial, and imbued with assumptions and politics which can be contested \cite{haraway2004situated, latour_drawing_1990}. A case in point is the Mercator cartographic projection which inflates continents near the poles at the expense of land masses near the equator. Another example is a bibliometric measure of the h-index which reduces the ``impact’’ of a researcher based solely on the ratio between the quantity of publications and citations.

In contending with the above critiques of classification and visualization, some researchers and practitioners have been interested in exploring how to democratize knowledge production through more inclusive and inter/trans-disciplinary collaborations \cite{gibbons2001potential, o2022capabilities, wilson2017new}. Theoretically, a significant body of research has been motivated by the question ``how do diverse actors create a common understanding without losing the identity and autonomy of their social worlds?’’ \cite{star1989institutional, trompette2009revisiting, carlile2002pragmatic, michalec2021reconfiguring, stoytcheva2013boundary}. Here, the concept of ``boundary objects’’ has been particularly useful for understanding the dynamics of collaboration. The term originates from the foundational work by Star and Griesmer \cite{star1989institutional}, and since has been commonly adopted across the CSCW literature, see e.g., \cite{bodker1994scenarios} on scenarios in design; \cite{cai2021onboarding} on onboarding materials, or \cite{bossen2014boundary} on healthcare records. In short, boundary objects are keywords, documents and artifacts that allow diverse groups to work together without consensus \cite{star1989institutional}. The key features of boundary objects are their interpretive flexibility, diverse structures of information needs, and, finally, the dynamic between ill-structured and more tailored uses of the objects \cite{star2010not}. Importantly, boundary objects do not reflect ``things out there’’, rather they derive from an intention to collaborate and achieve common goals. Wenger \cite{wenger2000communities} outlines activities necessary for successful collaboration at the boundaries of expertise: a) Abstraction facilitating a dialogue between communities of practice; b) Multi-tasking: several activities or practices are possible with a single boundary object; c) Modularity: different parts of the boundary object can serve as a basis for dialogue between actors; d) Efforts towards standardization of the information contained in the object to render the information interpretable. 

Collaboration, however, does not always guarantee more democratic or inclusive outcomes. Issues like institutional inertia and a lack of capabilities to maintain networks over time trouble the attempts of creating knowledge across siloes \cite{huxham2004doing, o2022capabilities}. Collaborative approaches in research (often called co-design, co-creation or participatory research) have been criticized for the lack of conceptual clarity, the tensions they create between the open-ended nature of creative work and the requirement to tailor research proposals at an early stage, time pressures, expectations of impact, tokenism and epistemic burden, or insufficient resourcing and experience from community stakeholders \cite{oliver2019dark, pierre_getting_2021}. Yet despite their numerous challenges, studies also show a promising path and growing demand for research involving co-design and participatory approaches. This is particularly the case in addressing complex issues of technology ethics, harm and injustice~\cite{birhane_power_2022, dencik_data_2022, costanza-chock_design_2020}. For example, recent CSCW research on the participatory classification of online harassment \cite{blackwell2017classification} argued that fully addressing online harassment requires an ongoing integration of vulnerable users' needs into the design and moderation of digital platforms. Similarly, research on participatory threat modeling encouraged traditionally marginalized people to define their own cyber security threats and preferred defense measures~\cite{slupska2021participatory, das_chowdhury_utility_nodate}. Advances in participatory methodologies have also extended to visualization, where creative techniques have been used to facilitate and illustrate conversations centered around the lay users’ experiences of computers and insecurity \cite{heath2022drawing}. One of the main achievements of this strand of work has been a critical return to the notions of positionality and expertise, i.e., questioning who gets to frame, work on or benefit from research and classification activities. Knowledge infrastructures, if created in a collaborative way, tend to prioritize open access, continuous editorial process, and experimentation with regard to visual communication \cite{aspesi2020pursuit}. Collaboration also opens opportunities for productive disagreement, as stakeholders are actively encouraged to deliberate over their opinions in a structured and facilitated format.

Building on this agenda, the mapping process and products we describe here, can be best understood and advanced through the lens of collaborative knowledge infrastructures and co-design. In much the same way as boundary objects, our online harms map is intended to be a gathering point between different communities not only for hosting academic literature, gathering policy evidence and scanning the research landscape but also for encouraging multi-stakeholder collaboration and dialogue beyond the academy. In the next subsection, we review extant efforts to define, classify and taxonomize online harms within different academic communities. 

\subsection{Theorizing, taxonomizing and sorting online harm}

There is a vast body of research concerned with individual and social harms linked with the use of the internet and digital technology at large. The phrase ``online harms'' has more recently been used in academic and policy literature as a shorthand, perhaps more so in Europe following the publication of the UK government’s Online Safety Bill. In this context,the Online Safety Bill defines ``online harms" as ``user-generated content or behavior that is illegal or could cause significant physical or psychological harm to a person"~\cite{dcms_research_2019, DCMS2022safety}. We note that while we use the phrase in this paper due to its increasingly common usage in some academic, policy, and practitioner communities, we do not endorse the above definition and in fact flag its conceptual limitations. For instance, said definition focuses on ``user-generated content or behavior" in an exceptionalist way while under-defining the role of institutional actors as well as other collective or social harms (e.g., harms to democracy). But there exist several other idioms referring to cognate and overlapping issues, some examples are data harms, online abuse, or cyber threats, risks, and vulnerabilities. Further, several subfields have emerged or built upon previous research in response to ethical concerns of information technology which are themselves adjacent to questions of online harm; some of these include, inter alia, data ethics, computer ethics, AI ethics, and responsible innovation. While we do not review the literature here (see \cite{cork_rethinking_2022} for a systematic review), it is pertinent for our purposes to make some broad observations. 

Because of the complexity and multiplicity of these topics of research, numerous schematizations and taxonomies of online harms, risks, and vulnerabilities have been borne out of diverse disciplines. Depending on their specific aims, these efforts seek to advance conceptual understanding, systematize empirical evidence, develop interventions, or inform policy around online harm. As noted by Cork and colleagues' \cite{cork_rethinking_2022} recent review, taxonomies from computational and information science disciplines tend to be broadly concerned with detecting and mitigating harmful content or cyber threats through different data-driven techniques (e.g.,~\cite{banko_unified_2020}), whereas taxonomies developed from social policy or social science disciplines tend to be primarily concerned with how best to define, evidence and conceptualize different types of harms (e.g., ~\cite{livingstone_risks_2010, scheuerman_framework_2021, blank_benefits_2018}), or inform the legislation of privacy and internet related harms (e.g., \cite{citron_privacy_2022, clark_toward_2019}).

Depending on their specific aims, online harm taxonomies offer different approaches to distinguish between the ``types'' of harm that exist. While technical taxonomies of online harm often focus on the specific factors which can lead to harm---such as technical vulnerabilities ~\cite{garg_analysis_2019}, perpetrator intentions~\cite{ghadge_managing_2019}, or methods used to inflict harm~\cite{brar_cybercrimes_2018}---social science taxonomies foreground broader social impacts or dimensions of harm e.g.,~\cite{agrafiotis_taxonomy_2018, livingstone_risky_2008,oconnell_young_2006}. For example, Livingstone et al. ~\cite{livingstone_risky_2008} propose four general ``motivations'' of online harm---aggressive, sexual, value-based, and commercial harms,  whereas O’Connell and Bryce~\cite{oconnell_young_2006} suggest five ``themes'' of harm---information, human interaction, health/body/spirit, sex education/recreation and communication, and activities harms. 

The notion of harms associated with digital technology has already received considerable attention within the CSCW scholarship, even if not explicitly under the rubric of online harm. For example, recent papers have applied frameworks from mental health research to discuss ``digital self-harm'' in the context of eating disorders as well as the correlation between harmful events offline and online \cite{pater2017defining,pater2019exploring}. CSCW research has also taken interest in harm reduction through the provision of safe spaces online, e.g. for queer communities intending to come out or for transgender people to explore their identity \cite{dym2019coming, scheuerman2018safe}. Another major theme of research is an exploration of online harassment experiences and the provision of moderation guidelines; with key contributions emphasizing the need to integrate vulnerable users into the co-design of recommendations and prototypes \cite{blackwell2017classification, blackwell2019harassment}. It is worth highlighting that CSCW has a long history of research defining, measuring, understanding, and tackling discrimination and abuse online without adopting the terminology of harm, see foundational papers on racism, justice, and bias \cite{hutson2018debiasing, fox2017imagining, lee2019procedural}.  

All in all, the landscape of research on online harm is marked by a diversity of research agendas and a lack of common vocabularies and definitive boundaries. These complexities pose numerous challenges regarding collaboration, particularly among scholars who are committed to different research paradigms, goals and methodologies, and who may disagree on concepts or interventions. A salient example of an ongoing debate is the concern by privacy advocates that tackling Child and Sexual Abuse Materials (CSAM) by weakening provisions for end-to-end encryption could legitimize more surveillance by the State or technology companies~\cite{ludvigsen_yasm_2022, russon_should_2021}. 

These challenges and tensions were part and parcel of our own attempt at building a collaborative knowledge infrastructure intended to map the terrain of ``online harms'' at the confluence of some of the disciplines mentioned here. While a review of the literature was a key input to the process, the goal of the map was not to develop a comprehensive inventory of harms or a static taxonomy, but a usable, configurable, and maintainable knowledge infrastructure. 

\section{Co-creating a map of online harms} 

\subsection{Conception and rationale}
\label{sub:conception}

The REPHRAIN Center is a major interdisciplinary community focusing on investigating, reducing and tackling online harm. The Center was funded by UK Research and Innovation in the context of a national policy agenda around online safety. It gathers over 100 internationally leading experts from academic institutions working across 37 diverse research projects and 23 founding industry, non-profit, government, law, regulation and international research Center partners. The Center works collaboratively across disciplines on a variety of issues around privacy, security, data sharing in the digital economy, content moderation and technology-mediated harm.

In addition to funding individual research projects, the Center employs ``core researchers'' (ADH, KMR, PDC, OM) who work on cross-cutting issues pertaining to the online harms landscape aiming to facilitate transdisciplinary work between projects, conduct scoping and horizon scanning work, integrate responsible innovation, engage policymakers, and raise the profile of the Center to external stakeholders to boost its impact and visibility. Alongside a team of core researchers, author (AC) worked on a project on defining and quantifying the notion of ``online harm'', while authors (EJ, EG) were employed as research assistants reviewing and cataloging the outputs of the Center at large. 

A key outcome of the Center---the ``map of harms''---was envisioned at the outset as a living, interactive, resource to showcase ongoing research within the Center as well as identified research gaps, relevant literature, and useful research tools and materials linked to particular themes\footnote{A live version of the map can be found on https://www.rephrain.ac.uk/rephrain-map/}. The specific format and affordances of the map were not decided a priori. Unlike the research projects funded under the Center which had a defined deadline, methodology, disciplinary orientation, and resources, the map was loosely defined and managed by the core researchers who led the co-design process in an iterative and experimental fashion. Broadly, the map was conceived with the following long-term aims in mind:
\begin{itemize}
    \item to facilitate the communication of research findings and policy recommendations to different stakeholders within and outside academia;
    \item to boost the profiles of the researchers affiliated with the Center;
    \item to help scope the future funding agenda, as aligned with identified research gaps. 
\end{itemize}

As part of the Center's bid for funding, a preliminary list of harms (see Table~\ref{table:OrigionalListOfHarms}) was developed drawing from two sources:  Daniel Solove's taxonomy of privacy paper~\cite{solove_taxonomy_2006} and the UK government's online harms white paper~\cite{dcms_research_2019}. This list informed the funding of projects and the agenda of work with external partners, and served as the starting point for discussions concerning the map.

\begin{table*}[!htp]
\centering
\small
\caption{Original list of online harms}
\renewcommand{\arraystretch}{1.2}
\begin{tabular}{l l l}
\hline
\hline
Data Availability & Child Protection & Disinformation \\ 
Pornography & Surveillance \& Government Incursion & Harassment \\
Inappropriate Disclosure & Identification & Intimidation \\
Consensual Mechanisms & Lack of Protection & Hate Crime \\
Terrorist/Extremist Content & Violent Content	&  Information Probing \\
Sale of Illegal Goods &&\\
\hline
\end{tabular}
\label{table:OrigionalListOfHarms}
\end{table*}

\begin{figure}[!htp]
         \centering
         \includegraphics[width=0.98\textwidth]{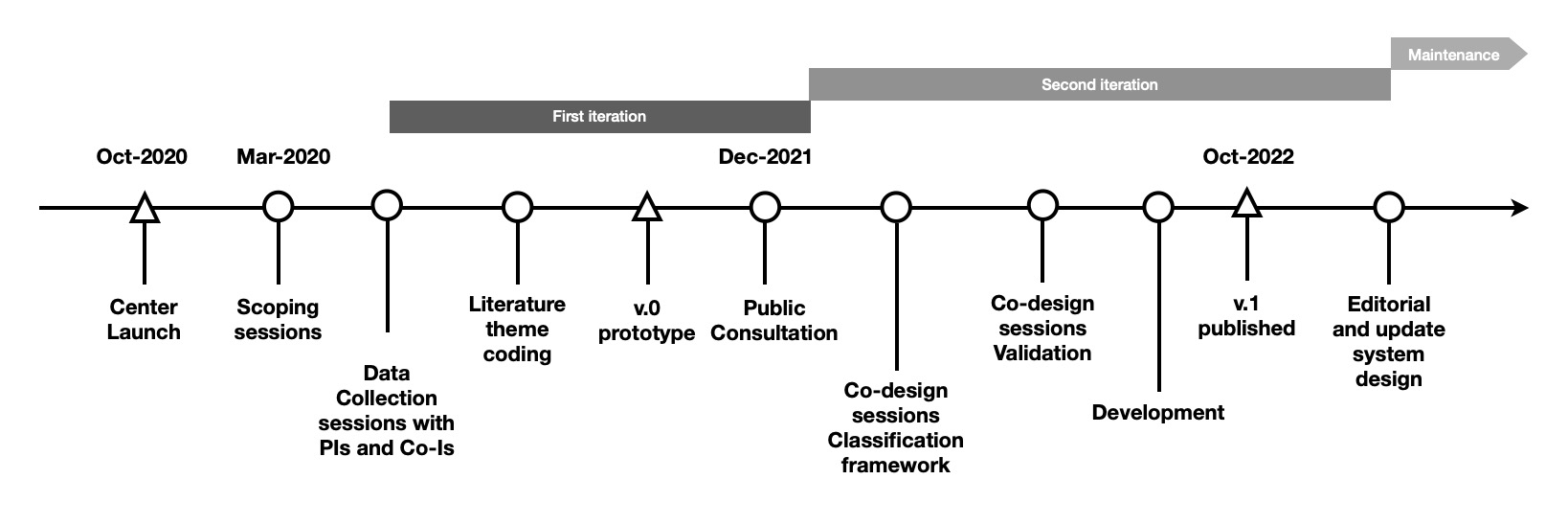}
         \caption{A timeline of the map co-creation process}
         \label{fig:maptime}
\end{figure}

In the following, we provide an account of how the map was brought to life and what utility it offered to different actors. We argue that the map---and the process of mapping---perform three mutually constitutive functions, acting simultaneously as a method, medium, and provocation. We ground our analysis on an 18-months process of co-creating a map of harms in collaboration with around 75 investigators and partners associated with the Center (see Figure \ref{fig:maptime}). We reflect on the use of different methods of data collection and curation, standardization, collective deliberation, prototyping, synthesis, and design that contributed to the construction of the map. The co-creation process was led by the team of core researchers from different academic backgrounds (computer science, human-computer interaction and social sciences) who organized the data collection and coordinated activities with the Center stakeholders. The iterations of the map were discussed and validated regularly at Center-wide events and with the Center's leadership in strategic meetings. Throughout this period, the core team conducted a series of workshops, one-on-one interviews, online surveys, public consultations, and design sessions in collaboration with various stakeholders including mainly academics affiliated with the Center, but also industry partners, policymakers, and members of the public. Given that the map is, to the best of our knowledge, the first of its kind, the process did not follow a predictable, linear trajectory but was instead informed by an iterative and trial and error approach. The following analysis is therefore not chronological nor intended to provide a template of best practice. We draw lessons from the activities performed in order to foreground the different dimensions of collaboration in this arena as well as the mutually shaping interplay between each of the three map functions. 

\subsection{Map as method}
\label{sub:method}

During the launch of the Center, the core team held a series of exploratory scoping sessions aimed at identifying various types of online harm and how the Center may address them. These sessions sought to elicit views of various communities as well as inform the design of subsequent activities and the questions to be explored in them. We invited participants from existing networks of academics who had converging interests with the Center's research and those involved in its conception to discuss what online harm is and what they would expect from a map of online harms. The scoping workshops (9 in total) were attended by 40 participants (21 academics and 19 representatives from academia, industry, law enforcement, safety tech developers and policymakers. The discussions focused mainly on what counts as privacy and other related harms; what approaches, tools and methods exist to mitigate these harms; the potential misuses or malfunction of technical interventions; and how such failures could be prevented while providing adequate protections.

These workshops called for more evidence and discussions around the prevalence of online harm, which harms are emerging or are yet to be addressed in the literature, their impact on different individuals and communities, what are the approaches and tools to mitigate harms, and open research, technical and regulatory challenges. They also highlighted challenges around addressing or reducing online harm in various spaces, sectors, situations, and organizations. Lastly, discussions around the map raised questions around what the map should offer, what features are critical, what audiences should the map target, as well as how the map should look.~\footnote{See \cite{ramokopane_scope_2021} for a detailed report} 
%The findings of these workshops were summarised and published as a report [anonymized citation].

Throughout the different stages of mapping, and particularly during the exploratory stages, positioning the map on the horizon helped not only to scope pertinent questions for its development, but also set the scene for deeper discussions about terminology, concepts, interventions, and research methods. We found that the concept of the map was useful as a dialogical tool that enabled researchers to link up different bodies of knowledge, access research from other disciplines and translate concepts from discipline-specific jargon. Interim sketches and depictions of the map were helpful to spark discussions about how can we best visualize a complex arena of research and what are the implications of such representations. 

\begin{figure*}
    \centering
    \includegraphics[width=0.5\textwidth]{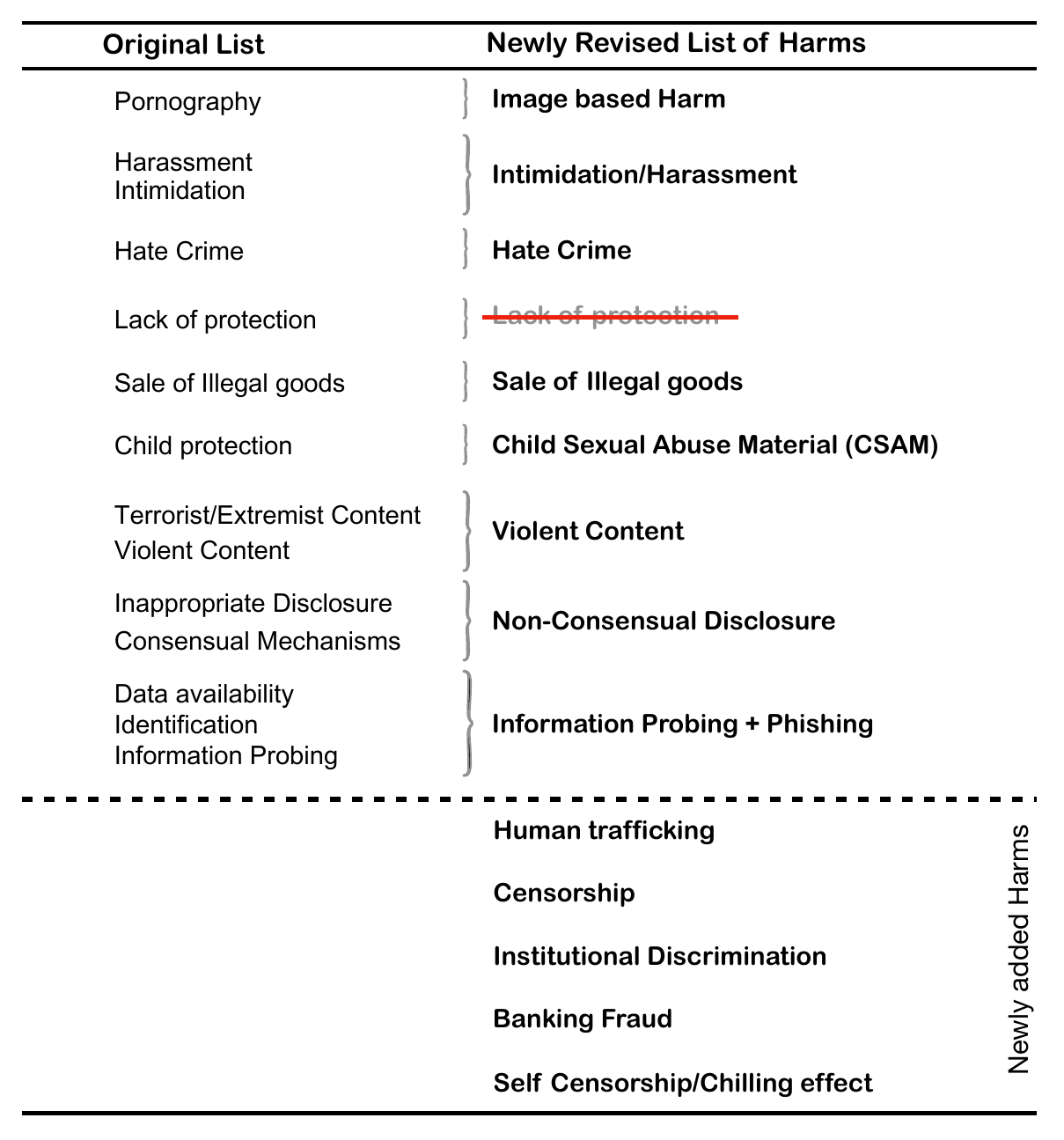}
    \caption{Revised list of online harms}
    \label{fig:CurrentListOfHarms}  
\end{figure*}

In order to materialize the first iteration of the map, we organized data collection sessions targeted at individual projects within the Center. We asked project Principal investigators (PIs) and Co-investigators (Co-Is) to complete an online form detailing (1) what harm(s) were being addressed by their projects, (2) a brief definition/description of the harm, (3) a list of research gaps, challenges or questions in relation to the identified harm(s), (4) the current state of the art including peer-reviewed academic articles, policy documents, white papers and reports, and lastly, (5) the technical, conceptual or methodological tools (both internally developed or external) to study, understand and addressing such harms. These responses were later used during face-to-face meetings to prompt investigators to expand or clarify their responses and how they could be accommodated in the form of a map. %Materials and data collected during these sessions include harm definitions, identification of research gaps, challenges and questions, external tools, datasets and resources related to online harms, relevant literature including peer-reviewed articles, policy documents, white papers and reports, and an inventory of expected deliverables by projects within the Center.
These data gathering sessions crucially helped to refine the initial list of online harms in terms of the adequacy of the terminology used (for example revising ``pornography'' for ``image-based harm''), and they revealed a need to use lay and concise descriptions as well as add, remove or merge harms in accordance with ongoing work within the Center and the state of the academic debate (see Figure~\ref{fig:CurrentListOfHarms}). This process of expert consultation, albeit relatively slow,~\footnote{Data gathering meetings were onerous for both core researchers and project investigators with only a few meetings conducted per month.} was key to help the core-researchers curate and organise data in areas outside of their specialties. In this sense the core-researchers deferred to the project investigators to provide authoritative content yet without foreclosing further modifications and inputs from other stakeholders. 

A parallel, and asynchronous, data curation process was conducted with the aim of further populating the map with relevant and up-to-date literature produced by those whose work was expected to feed into the map, but might not have yet been approached through face-to-face data collection meetings. To do so, publications by all Center-affiliated researchers were manually collected, filtered, theme-coded and mapped onto the evolving list of harms. First, EJ and EG manually screened the titles and abstracts of a total of 232 papers for whether they addressed one or more harms within our list~\footnote{To ensure the review of publications was as broad as possible, papers which considered online harms from the Center’s initial list as well as harms which were not listed by the Center at the time of screening were included. Papers which did not address or specify harms related to online activities or platforms were excluded.}. Then, a closer reading and depuration of the remaining papers were done to sieve out those that did not refer to or address online harm or the topic in any form. This process led to 125 papers being included in the first iteration of the map (see Table \ref{tab:samplecodes}. 

\begin{table}[!ht]
    \footnotesize
    \centering
        \caption{A sample of codes extracted from research papers. (IPV stands for intimate partner violence)}
    \begin{tabular}{l|l|l|l}
    \hline
    \hline
        \textbf{Victims } & \textbf{Perpetrator } & \textbf{Platform/Technology} & \textbf{Methodology} \\ \hline
        Bystanders  & Campaign groups  & AI systems & Anomaly detection \\ 
        Children  & Darknet communities  & Internet of Things & Case studies \\ 
        Consumers  & Extremist groups  & Cloud systems & Usability Studies \\ 
        IPV victims  & Government agencies  & Contact tracing apps & Detection system \\ 
        Online dating users  & IPV perpetrators  & Content-sharing
                                                services & Digital forensics \\ 
        Political organisations  & Law enforcement  & Critical infrastructure & Digital traces \\ 
        Sex workers  & Nation State  & Darknet markets & Ethnography  \\ 
        Social media users  & Online fitness  communities  & Emails & Experimental \\ 
        Teenagers  & Organized crime groups  & E-recruitment platforms & Focus groups \\ 
        Refugees & Romance Scammers & Virtual Reality & Interviews \\ 
        Bystanders & Sex Offenders & Social Media Platforms & Surveys \\ 
        Women & Social Media Users & Smartphones & Social Network Analysis \\ \hline
    \end{tabular}
    \label{tab:samplecodes}
\end{table}

The literature curation exercise offered a useful overview of the diversity of expertise and disciplines across the Center including, e.g., technical approaches to harm mitigation, methods for measuring or gathering evidence, policy interventions, or social scientific approaches to understanding harm. These papers were theme-coded according to the harm(s) they addressed and five high-level positive categories (developed from further discussion sessions, we discuss these in section \ref{sub:procovation} ): privacy, safety and well-being, reputation, financial security, freedom of speech, and fairness. Although these methods of bibliographic analysis evoke (and could well lead to) a formal systematization of knowledge or literature review, that was not the primary aim. Instead, this exercise sought to be continuous and directly functional to the map: to showcase the online harms work within the Center in an useful way to different stakeholders and in relation to the evolving classification affordances of the map.

In aiming to improve the functionality and usability of the map, and responding to feedback from a community consultation (see section \ref{sub:medium}), the team decided to further classify papers according to their methodologies (e.g., case studies, focus groups, or interviews), the type of victim (e.g., children, teenagers, sex workers, women), the type of perpetrator (e.g., romance scammers, extremist groups, sex offenders), and the technology or platform being studied (e.g., artificial intelligence systems, virtual reality, social media platforms). While these new categories emerged from a limited set of papers and are therefore not exhaustive, the expectation was that more granular information would offer users more options to navigate the map or find interconnections (or lack of) between different papers, authors, harms, technologies and attributes. Similarly, while the goal was to construct a visual representation of literature on harm, curating our collaborators' input and theme-coding the various attributes of harm in different ways led us to devise methods of synchronous and asynchronous collaboration, synthesizing previously disperse bodies of knowledge, and conducting meta-analysis in ways that were unexpected, and yet now standardized thanks to the use of codes and tags.
As we will show next, this aspect of knowledge categorization was a key input for shaping the content and structure of the information presented to users. 

\subsection{Map as medium}
\label{sub:medium}

%Target groups:  academia, industry, and law enforcement, Safety Tech developers, and policymakers.
Translating our collaborative work into various inscriptions, diagrams, mind maps, schematics, and sketches ~\cite{latour_drawing_1990} was a necessary endeavor in envisioning and materializing the map. Partial and interim depictions were used not only as milestones of progress toward fulfilling the intended knowledge-sharing function of the map, but as useful in unexpected and practical ways: containing a body of knowledge and definitions that informed others' research and literature reviews, linkages between researchers and pointers to their academic profiles, curated lists of papers, and areas of harm where more attention is needed. 

\begin{figure*}[!h]
     \centering
         \centering
          \includegraphics[scale=0.18]{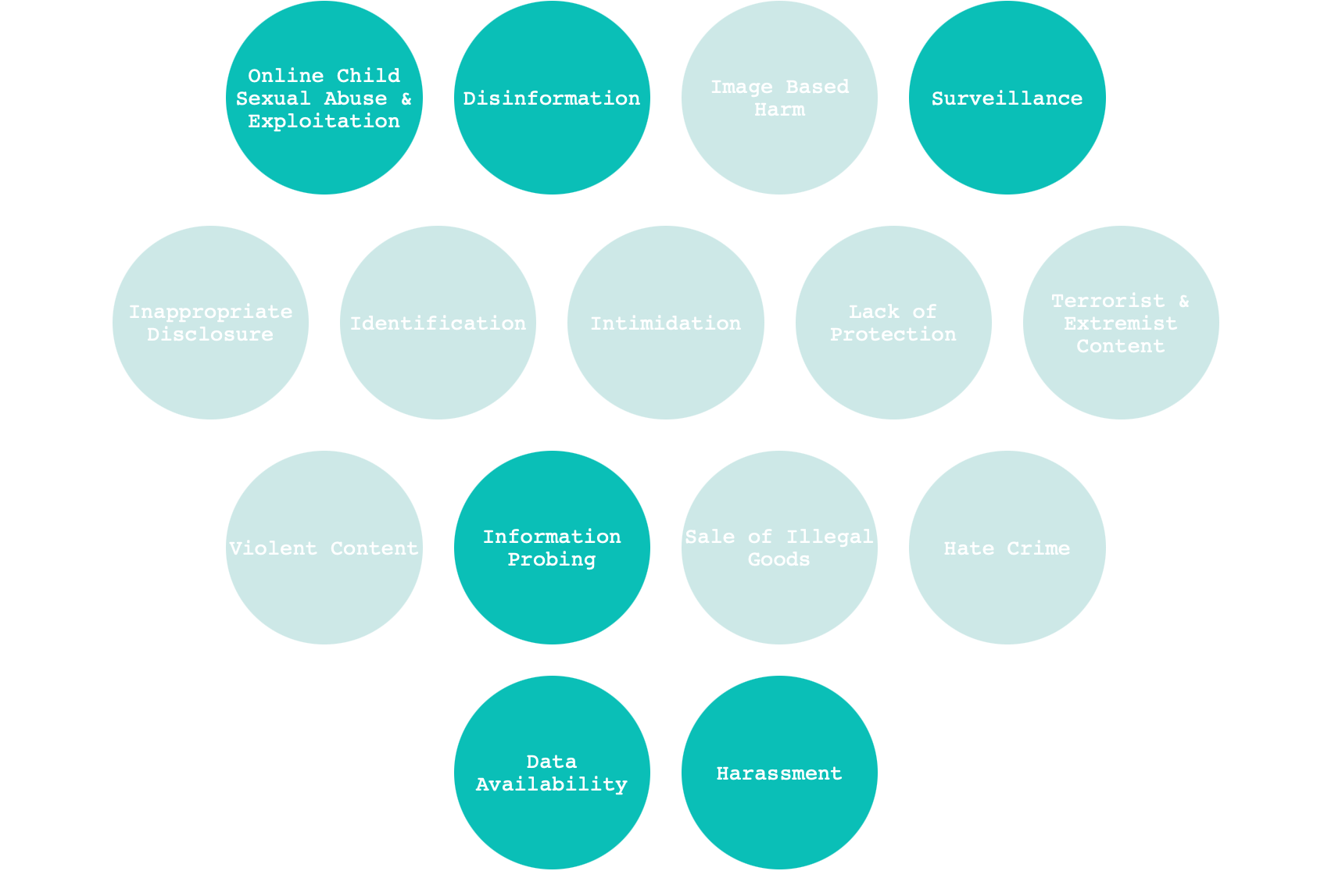}
   \caption{The first prototype of the map (v.0) used for community consultation}
   \label{fig:map_v0}
     \hfill
\end{figure*}

With the data collected from projects, we developed a prototype of the map which included a visual representation of the Center's list of harms, each of which would be populated with information containing definitions; identified research gaps, challenges and questions; external tools, datasets and resources related to online harms; relevant literature including peer-reviewed articles, policy documents, white papers and reports; and an inventory of expected deliverables by the Center projects. This first prototype---known as v.0 (Figure \ref{fig:map_v0})---aimed not only to offer a visual aid for further data collection sessions by showing placeholders where data is required, but crucially to prompt further discussions about the role of the map, how it should look, its intended users and its implications. The map prototype included 6 harms containing relevant information, definitions and resources, as well as placeholders (in the form of greyed-out circles) for harms where data were still needed. 

\begin{figure}[!htp]
         \centering
         \includegraphics[width=0.9\textwidth]{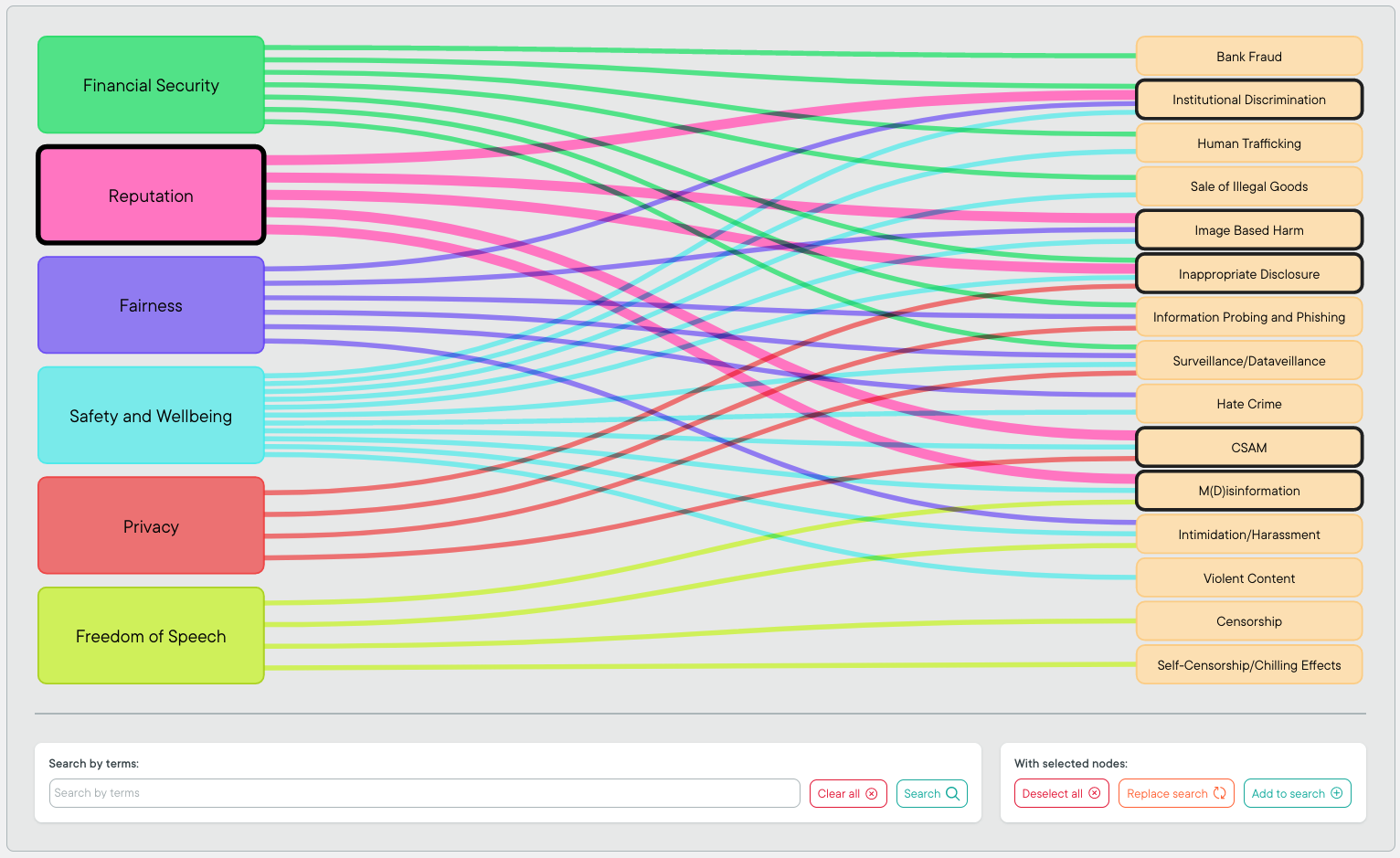}
         \caption{First public release (v1.0) of the map}
         \label{fig:map_current}
\end{figure}

\begin{figure}[!htp]
         \centering
         \includegraphics[width=0.9\textwidth]{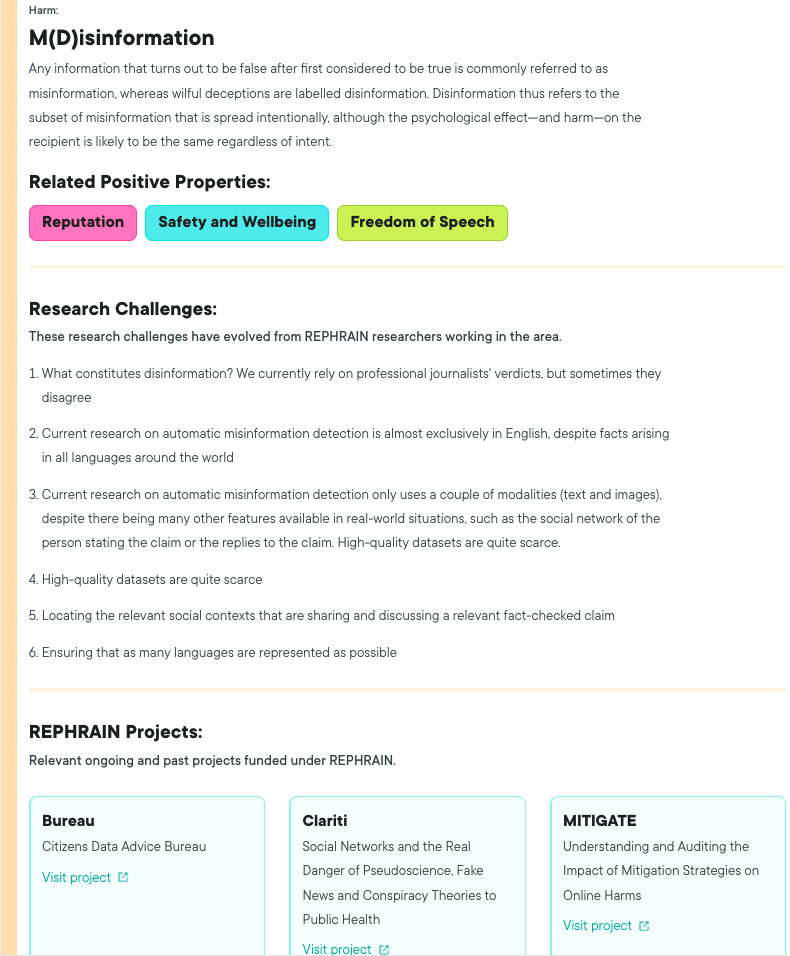}
         \caption{Example: M(D)isinformation page}
         \label{fig:disinformation}
\end{figure}

\begin{figure}[!htp]
         \centering
         \includegraphics[width=0.9\textwidth]{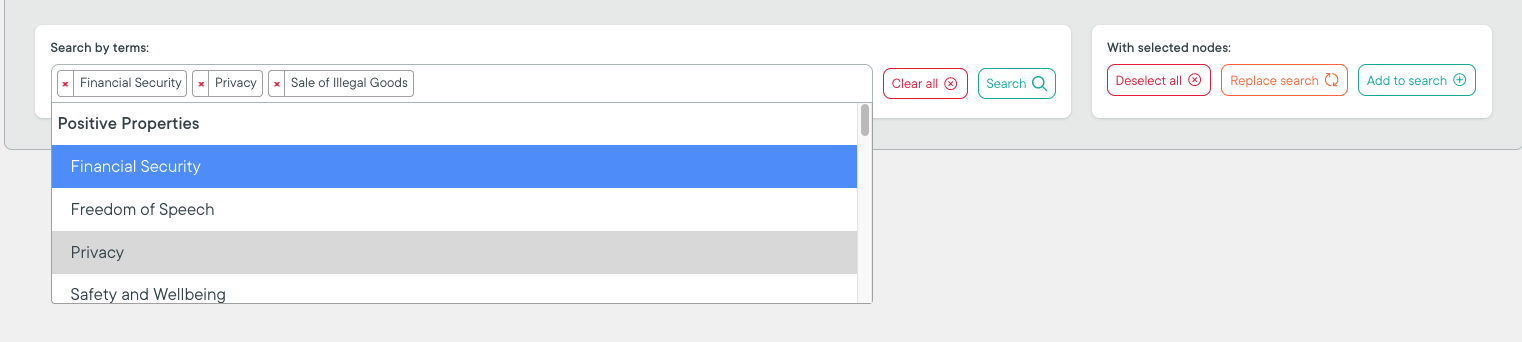}
         \caption{Search functions}
         \label{fig:mapfunctions}
\end{figure}

\begin{figure}[!htp]
         \centering
         \includegraphics[width=0.9\textwidth]{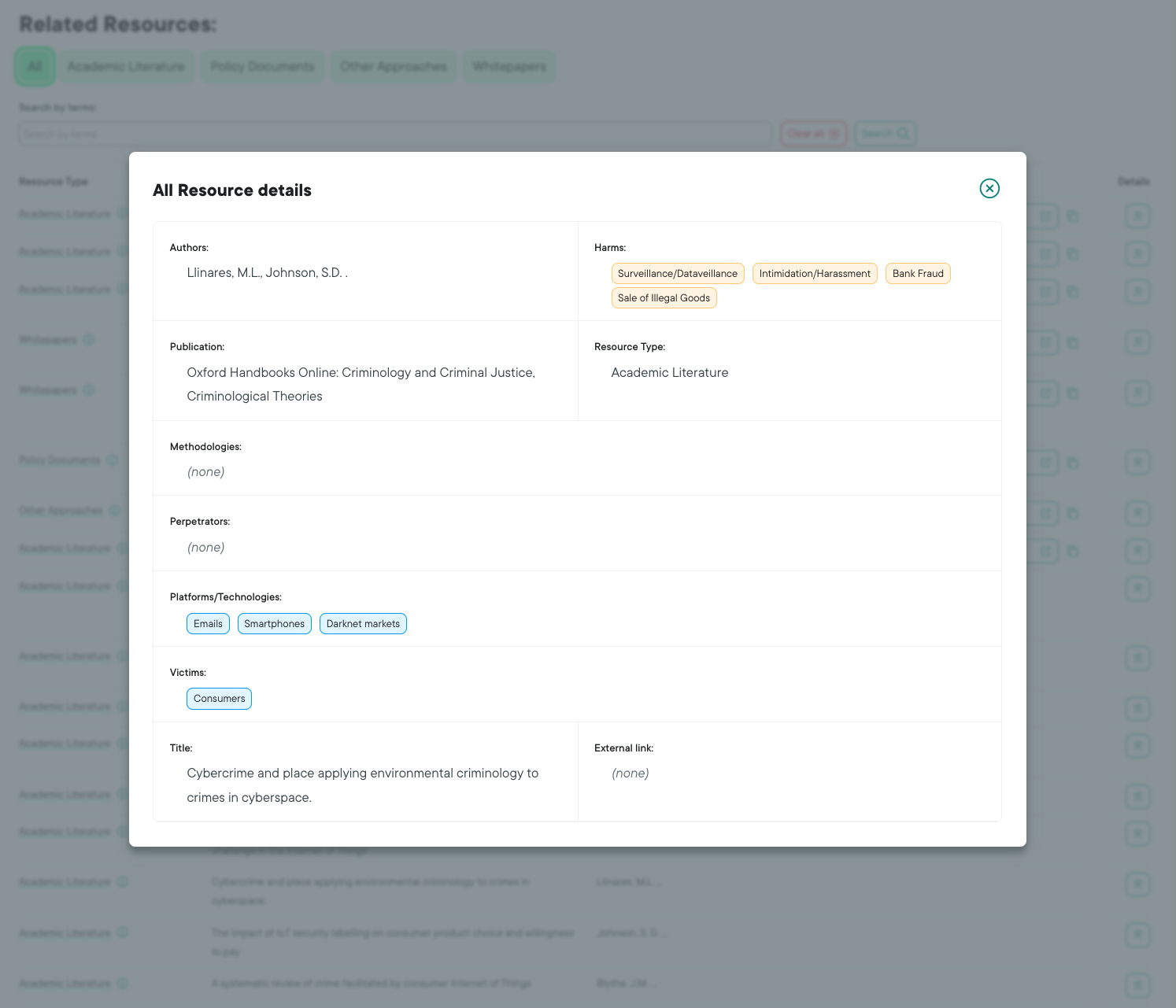}
         \caption{Theme coding functions}
         \label{fig:mapfunctions2}
\end{figure}
After the launch of the first v.0 prototype, a public consultation was conducted seeking feedback on different aspects of the map such as look and feel, content, technical features, and other open-ended suggestions. We disseminated a link to the map prototype and an online questionnaire through various communication channels including mailing lists of allied networks of academics and social media (Twitter and LinkedIn). This was the first time the map was shared publicly to external stakeholders from industry partners, third-sector and civil society organizations. We received 7 anonymous responses during the public consultation period between 27/Nov/2021 to 14/Dec/2021. The feedback and recommendations from the public consultation were analyzed in a project management platform, grouped and theme coded into the following: map structure, layout/look and feel, content and literature, definitions of harms, use cases and features, other modifications of the map and additional comments. Recommendations were also prioritized according to their feasibility to implement as ``immediately,'' ``in short term,'' ``in a second map iteration,'' ``for later re-consideration'' or ``no action.'' Some of the prioritized recommendations included the need for providing users with an explicit description of what the map aims to achieve; explanations about how the list of harms came about; and how the literature was selected for each harm. 

The public consultation highlighted important aspects about the usefulness, purpose and implications of the map. One of the core pieces of feedback was the need to be explicit about the purpose and target audience of the map. For example, it was suggested that while some academics may be interested in the theory pertaining to online harms, policymakers would be more concerned with actionable evidence, as well as understanding the practical implementation of harm reduction and mitigation strategies, while computer scientists may be interested in the technical challenges of combating or measuring online harm. In an attempt to address this issue, we proposed that by breaking papers down into their components of harm ---victims, technologies, methodologies, platforms--- users of the map could more easily access a very heterogeneous knowledge base. In this way, a modular and interactive approach could be the most practical, wherein the audience could personalize their experience of navigating the map and find relevant material (see section \ref{sub:method}). Breaking online harms down into their component parts could also allow users and curators to identify gaps and additional insights in the research being undertaken, for example, areas where certain types of victims may be underrepresented in the research, or identify parallels in mitigation approaches across multidisciplinary perspectives. 
%This components-based approach to mapping online harm also enables us to understand precisely how the context around particular cases of harm - e.g., misinformation or violent imagery, serves to increase vulnerability or resilience to harm. 

This learning and the potential actions being proposed were later discussed in co-design sessions involving the core team and a team of designers and web developers with the aim of creating a new iteration of the map. In these sessions, different forms of visualizing and filtering data were workshopped and trialed during a period of three months. Figures~\ref{fig:map_current}, \ref{fig:disinformation}, \ref{fig:mapfunctions} and \ref{fig:mapfunctions2} show the initial public release and the various search functionalities of the map. The maintainability and ``future proofing'' of the map were also considered at this stage: the back-end of the system was designed such that new data and edits to the current data would be conducted through GitHub.  %The next release of the map is shown in Figure~\ref{fig:Mapv1}. 

\subsection{Map as provocation}
\label{sub:procovation}

While the map progressed incrementally through discussion and (at least provisional) consensus between collaborators, it was also frequently a source of contention and disagreement. For instance, a recurring point of debate during co-design workshops was due to disagreement among researchers over the appropriateness of the term ``online harms'' for some of the issues addressed by the Center. This debate eventually led to the proposal to broaden the scope from simply a ``map of online harms'' to a ``map of technology-mediated harms, risks and vulnerabilities'' in order to encompass issues that cannot strictly be conceptualized as harms in their own right but could nonetheless lead to, or be linked with, harm (e.g., surveillance or misinformation). 

Much of this debate was informed by ongoing research on harm within the Center by AC% review of literature on harm conducted by Cork et al.~\cite{cork_rethinking_2022}. 
which found that harm and risk are both ill-defined in terms of causes and outcomes. Risk is often defined as a factor that may cause harm, but the risk is identified post-hoc---after harm has occurred. Risk is also thought of as the \emph{potential} for something to happen ---be that positive or negative. Harm is also subjective---both when experienced at an individual level and when thought of in relation to social values being harmed. By contrast, issues relating to abuse often involves a perpetrator and a level of intentionality which are not intrinsic features of risk and harm. Another conceptual challenge was to define ``technology-mediated'' or whether the idea of online harm stems from presumed causal relations between technology (or specific technology affordances, features, platforms, systems, business models) and harm. The modifier ``online'' was in this sense contentious in that it raised questions about the specific nature of harm being addressed by the map, the implication of a hierarchy of harms that could privilege online vs offline, and what harms might and might not be construed as pertinent in this academic program as a result. 

Several co-design workshops were organized where researchers and participants of all-hands meetings (quarterly center-wide events to showcase progress and discuss the strategic direction of the Center) were invited to provide feedback on the map and validate its iterations. These workshops aimed to explore different intuitive visual interfaces and ways to sort, categorize, visualize, cross-reference and represent information on the map. Inputs were sought from participants in the form of design sketches, recommendations to group/add/revise harms, and relevant factors which could help filter information and navigate the map. One of the key outcomes of these deliberations was the need to improve the visualization and grouping of harms, risks, and vulnerabilities such that they provide more useful information to users, and if possible, show the links between them. Rather than merely a matter of usability and aesthetics, linking and grouping items raised key issues about what knowledge claims are advanced by the map and what are their implications for different users. 

These issues were explored in an internal co-design workshop (among the authors of this paper) where we asked ``what are the harms, risks, and vulnerabilities we are studying a threat to?'' and ``What are the interventions we design aiming to protect or guarantee?''. In asking and answering these questions we drew inspiration from the threat modeling approach widely used in cybersecurity, as well as the United Nations human rights list. The result of this exercise was a framework for categorizing harms, risks, and vulnerabilities into five high-level positive categories or social goods, namely: privacy, safety and well-being, reputation, financial security, freedom of speech, and fairness. We mapped each of the harms, risks and vulnerabilities to one or more of the five positive categories as shown in Table~\ref{table:Framework}.

\begin{table*}[!htp]
%\tiny
\resizebox{\columnwidth}{!}{
\centering
\renewcommand{\arraystretch}{1.2}
\begin{tabular}{l l l l l l}
\hline
\textbf{Privacy} & \textbf{\makecell[l]{Safety}} & \textbf{\makecell[l]{Reputation}} & \textbf{\makecell[l]{Financial Security}} & \textbf{\makecell[l]{Freedom of Speech}} & \textbf{\makecell[l]{Fairness}}  \\
\hline
\hline
Surveillance/Dataveillance & Intimidation/Harassment	& Image Based Harm & Non-Consensual Disclosure & Censorship	& Institutional Discrimination \\
CSAM & Non-Consensual Disclosure	& Non-Consensual Disclosure & Surveillance & Self-Censorship/ Chilling Effects	& Intimidation/Harassment \\
Information Probing & CSAM	& CSAM & Human Trafficking & Intimidation/Harassment	& Image Based Harm \\
Non-Consensual Disclosure & Hate Crime	& (M)Disinformation & Sale of Illegal Goods & 	&  Hate Crime\\
 & Human Trafficking	& Institutional Discrimination & Information Probing & 	& Surveillance \\
& Surveillance & & Institutional Discrimination & & Information Probing \\
& Violent Content & & Bank Fraud & & \\
& Image Based Harm & &  & & \\
& Sale of Illegal Goods & &  & & \\
& Institutional Discrimination & &  & & \\
\hline
\hline
\end{tabular}}
\caption{ A framework for classifying harms against positive or desirable social goods}
\label{table:Framework}
\end{table*}

\begin{figure*}[!htp]
     \centering
     \begin{subfigure}[b]{0.32\textwidth}
         \centering
         \includegraphics[width=0.8\textwidth]{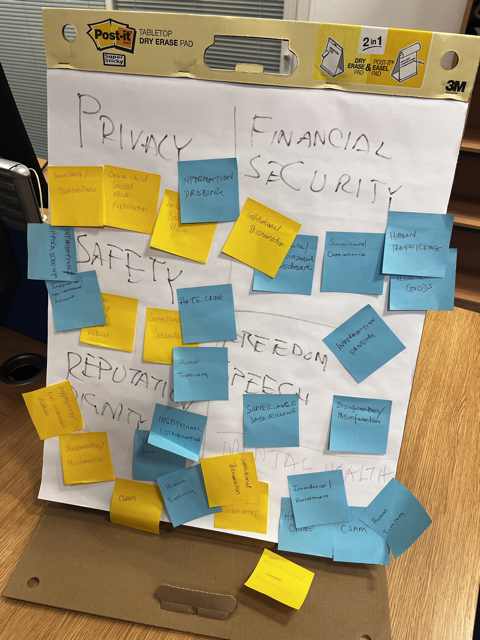}
         \caption{Group 1}
         \label{fig:group1}
     \end{subfigure}
     \hfill
     \begin{subfigure}[b]{0.32\textwidth}
         \centering
         \includegraphics[width=0.8\textwidth]{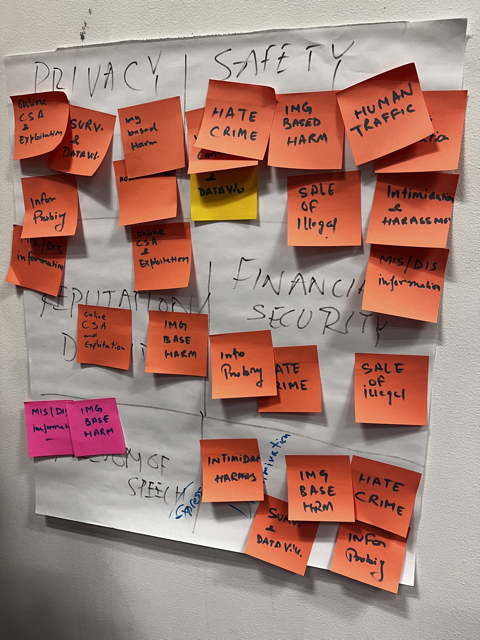}
         \caption{Group 2}
         \label{fig:group2}
     \end{subfigure}
             \hfill
    \begin{subfigure}[b]{0.32\textwidth}
         \centering
         \includegraphics[width=0.8\textwidth]{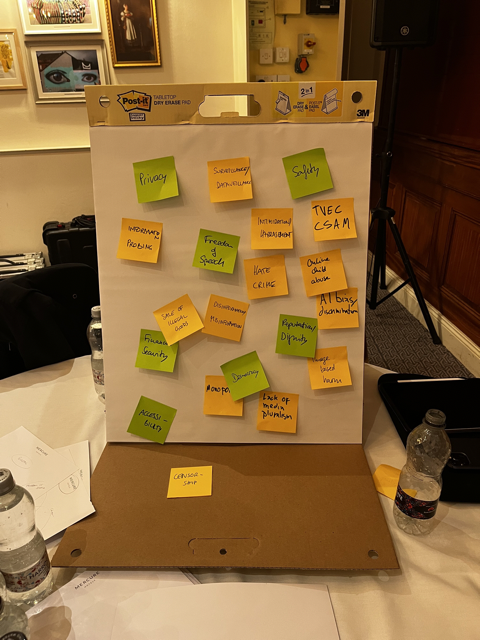}
         \caption{Group 3}
         \label{fig:group3}
     \end{subfigure}
        \caption{Work-shopping a framework to categorize harms}
    \label{fig:GroupActivity}
\end{figure*}

The framework was then validated in a follow up co-design workshop (for the record of the workshop process, see Figures \ref{fig:group1}, \ref{fig:group2}, \ref{fig:group3} ) as part of the Center's all-hands meeting where we asked participants to validate the utility of the categories. For this workshop, we asked participants working in three different groups (of 4 to 5 people each) to classify the list of harms against the positive categories or social goods using our suggestions as examples but not limited to them (i.e., harms could be mapped onto more than one category). We explained that while the five categories were not meant to be exhaustive they were intended to subsume the themes addressed by Center projects. Participants were encouraged to add new categories, remove irrelevant ones or change the terminology if needed. Similarly, participants were not bound to the suggested list of harms and were free to add new harms or refine the terminology. Each group then presented their results and rationale to the other groups. Using the collected material from each of the tables the core team analyzed and consolidated the results with updated terminology and classifications. The need for new categories and/or terminology would be evaluated as needed if new research did not fit the existing ones.

\emph{On Survivability and Maintenability}.
A key feature of the map, as envisioned in its conception, was its open-endedness and future relevancy. The ambition was to offer researchers the ability to update and refine the contents and structure of the map and ensure its survivability and maintainability beyond the lifespan of the Center, thereby prompting the need for a system for contributions and curation of new data entries. We discussed the design of such a system in all-hands meeting workshops with the Center researchers who raised questions of gate-keeping (who can contribute to the map and in what capacity?), frequency and types of contributions, technical requirements for updates and additions, and ongoing maintenance costs. 
These discussions led to the following conclusions: First, a streamlined system to populate the map should be used as a way to replace the manual and time-consuming process of requesting contributions from investigators via one-on-one meetings. Such a system should encourage contributions and cater to researchers with different technical skills and preferably rely on open-source software. Second, the system would require a data curation role to accept or reject contributions to populate and update the map, as well as contributor roles assigned to investigators within the Center and invited external contributors. Third, a workflow and other relevant documentation should be written to make the process transparent and guide decision-making and future updates to the map. Fourth, funds would need to be secured to ensure the map continues to be hosted, maintained and updated in the future. 

These goals remain open challenges at the time of writing of this article, not least due to ongoing debate among contributors with different technical skills about the choice of user interface for facilitating updates (e.g., wiki vs git), the policies to vet contributors of the map, and the availability of funds to maintain the project in the long term.

\section{Discussion: More than just a map}

The process of co-creating a map of online harms taught us valuable lessons about how knowledge representations emerge and how they get challenged or stabilized in transdisciplinary and inter-organizational collaborations. The purpose of the map was loosely defined from the outset with the expectation that all of the Center investigators and external contributors would help to shape and populate it. As we have shown, the map transcended its original scope and fixed temporality as a deliverable, not only serving as an open repository of knowledge about the range of research and the projects' outcomes, but allowing, throughout its construction, to uncover new insights and spark debate. The process of sourcing data and feedback from investigators opened up previously unforeseen challenges and opportunities, conceptual, methodological, and epistemic contentions or disagreements. It also highlighted competing views about the function and implications of the map. The process of mapping was by no means linear as interim findings, failures and moments of learning, importantly altered the initial goals and ambitions. Here, we have brought attention to three mutually constitutive functions performed by the map beyond its original aim as a deliverable. We demonstrated that our map of online harms simultaneously operates first as a ~\emph{method} of scientific collaboration, acting as a motive for dialogue between different communities and catalyzing modes of asynchronous cross-referencing of academic work. Second, it provides a ~\emph{medium} for knowledge representation and a repository that allows different stakeholders to sort out and find relevant information and a bird's eye view of multiple interconnections between harms, technologies, researchers, and disciplinary outputs. Finally, the map serves as a ~\emph{provocation} encouraging contention, dispute and disagreement, which in turn challenges the work of data curation in deciding the content, the form, the timing, the survivability, and the provenance of the knowledge that is represented in the map. 

These three functions necessarily inflect one another and render the map an always unfinished endeavor. In grappling with these facets of mapping an area of research that calls for sometimes urgent social and political action, the role of curators and facilitators are critical for dealing with the lack of consensus, recognizing provisional milestones, instigating and facilitating collaboration over time, taking pragmatic decisions and ensuring the maintainability of the map. By bringing attention to continuous iteration and feedback, we foreground the \emph{living} aspect of creating a knowledge infrastructure. Instead of treating the map of harms as a static, one off taxonomy, this effort shows the value of knowledge that gets updated, expanded, or even challenged within an existing network of collaborators. Ultimately, knowledge infrastructures---especially if pertaining to transdisciplinary and contested issues---display a ``rhizomatic'' character, that is, with multiple points of exit and entry and connected in multiple and surprising ways \cite{felix1987thousand}. Our experience of knowledge co-production reveals a networked, ambivalent and highly unpredictable process simultaneously opening space for the co-existence of plural and sometimes diverging views, and the evolution of ideas,  counterposed with the need for pragmatic utility, closure and standardized categories.

The emphasis on survivability evolved into design principles of modularity, customizability, transparency in the editorial policy, and concurrently an invitation to challenge the evolving knowledge base. Without it being the original purpose of the project, the emerging product evokes some of the affordances of knowledge management systems like Obsidian \footnote{https://obsidian.md/} or even Wikipedia \footnote{https://en.wikipedia.org/wiki/Main\_Page}, while still remaining restricted to a relatively small research community rather than aiming to be a universal taxonomy. 

Quite crucially, although our process of co-design has been generative in various ways, it is not without drawbacks. Despite the fact that the Center brings many experts together, the knowledge base is not intended to be exhaustive and indeed gives more visibility to emerging work, much of which from focused projects and early career researchers. So too, despite efforts to garner inputs from diverse stakeholders, the map was principally shaped by academic expertise and those knowledgeable of the Center but less so by non-academic groups and experts with lived experience of online harm. There are many practical reasons for this including the complexities with obtaining resources, ethical clearance and a strong case for involving external participants given the wide scope of the map in terms of domains of online harm. As a result the views from lay users and non-academic groups could only be indirectly represented by investigators engaging with such groups within their projects. The map is also not representative of all possible online harms but only those for which there is data and ongoing work in specific contexts and locations linked with the Center and allied collaborators elsewhere. Yet at the same time, there is an implied expectation of authoritativeness and generalized utility of the information it offers. These ambiguities and gaps are not easy to reconcile and might not be always transparent to users. As such, the map poses a challenge of communicating clearly the limitations and scope of the knowledge base without undermining its value to inform technology users and academics, the funding of further research, and policy and regulation. Another limitation is that the broad intended audience of the map risks addressing everyone and no one at the same time. The map might not meet the expectations of all its stakeholders, containing material that may only be useful to some, and that is biased towards the views, terminologies, mental models, and interface preferences of its curators and contributors. As much as the process of co-production has tried to be as open and inclusive without falling into the trap of endless debate, a challenge remains to enable the map to have useful entry points for various users and forms of expertise in the future. The success of the co-creation process is difficult to measure in this regard because there are no established benchmarks for evaluation and because the map can have intangible benefits as it is used and appropriated by different stakeholders in unexpected ways. This is an important question that calls for continual assessment in use.

Going forward, the future of the map remains open and we envisage multiple possible applications of the resource. First, recalling Star's theorization of the cycle of life of boundary objects \cite{star2010not, star1989institutional}, we anticipate that some of our stakeholders (e.g., our partners in civil service) would advocate for standardization and maintanance of (some of) the map content. For example, our knowledge base could be integrated into governmental documents and inform the work on around online safety and online harms regulation\cite{DCMS2022safety}. So too we hope that other non-academic users and contributors (e.g., civil rights organizations or victims of harm) can benefit from the open-ended format of the resource and input ideas from communities that are most vulnerable to online harms. Finally, the uncertainties associated with the research funding landscape put the long-term maintenance of the resource into question. Ultimately, this highlights the never finished and precarious nature of maintenance and curation, an issue deserving of care and appropriate funding in its own right \cite{ojala2022thesis}. 

\section{Conclusion}
In this paper, we reflect on the 18-month-long process of co-creating a knowledge infrastructure in the transdisciplinary context of online harms. We wish to bring to the fore the challenges of mapping an emerging body of work on highly contentious, unsettled, multifarious and pressing matters of concern. After an unstructured (and messy) co-design journey, we arrived at a malleable, collaborative and contestable resource that highlights several dimensions of technology-mediated harms, risks and vulnerabilities. Among other features, the map includes six desirable social goods, outstanding research challenges, signposting to foundational resources and researchers in each area, and modular filtering of resources.

Our contribution exemplifies how CSCW research could broaden discussions about transdiscplinary and inter-organizational collaborations to include useful reflections about the politics, discomforts, failures, pressures, residual prototypes, and lessons arising along the way in such co-productive efforts. By highlighting the three interrelated functions of the map (method, medium, and provocation), we were able to show the opportunities and challenges associated with collaborations across social worlds, the negotiation of boundary objects and the ambiguities of establishing an unfinished yet variously useful knowledge infrastructure. This is an important call for CSCW to foreground and engage with the multiple temporalities, social lives and political sensibilities of knowledge infrastructures.

\begin{acks}
This work was supported by REPHRAIN: The National Research Centre on Privacy, Harm Reduction and Adversarial Influence Online, under UKRI grant: EP/V011189/1. We would like to thank all contributors to the map and Yvonne Rigby and Robert Schultz-Graham for their support throughout the research.
\end{acks}

\bibliographystyle{ACM-Reference-Format}
\bibliography{main}

\end{document}